# Compact Acoustic Retroreflector Based on A Mirrored Luneburg Lens

Yangyang Fu[1,2,3,*], Junfei Li[1,*], Yangbo Xie[1,*], Chen Shen[1], Yadong Xu[3], Huanyang Chen[2,†]
and Steven A. Cummer[1,‡]

1. *Department of Electrical and Computer Engineering, Duke University, Durham, North Carolina 27708, USA*
2. *Institute of Electromagnetics and Acoustics and Department of Electronic Science, Xiamen University, Xiamen 361005, China.*
3. *College of Physics, Optoelectronics and Energy, Soochow University, No.1 Shizi Street, Suzhou 215006, China.*

We propose and demonstrate a compact acoustic retroreflector that reroutes probing signals back towards the source with minimal scattering. Gradient refractive index (GRIN) acoustic metamaterials, based on Archimedean spiral structures, are designed to fulfill the required refractive index profile. The experiments show that the compact acoustic retroreflector, whose radius is only approximately one wavelength, works in an incident angular range up to 120 degrees over a relatively broad bandwidth of about 27% of the central frequency. Such compact acoustic retroreflectors can be potentially applied in pulse-echo based acoustic detection and communication, such as unmanned aerial vehicle (UAV) sonar systems, robotic ranging detectors, ultrasonic imaging systems.

## I. INTRODUCTION

A retroreflector is a volumetric or surface structure that redirects the incident wave back towards the source, along the direction parallel but opposite to the direction of incidence. By rerouting the incident probing signal back towards the source or the transceiver, less signal is wasted in the uncontrolled scattering process and higher signal-to-noise (SNR) ratio of the sensing signal is obtained. In optics, conventional optical retroreflectors, realized with corner reflector or cat's eye configurations, can work within limited incident angles. Alternatively, by employing transformation optics [1-3] to transmutate a refractive index singularity, an Eaton lenses based on metamaterials [4-6] can be used as retroreflector [7] to broaden the limited incident angle range. However, the refractive index requirement for the Eaton lens based retroreflectors is extremely high. The resulting structures may have complicated geometries and high losses, which inevitably hinder their real applications.

To overcome these difficulties, we propose in this paper the use of Luneburg lens [8-10] with gradient-index profile and a rigid back layer. Different from ray-based designs, the gradient index wave manipulation allows for a more compact realization of the retroreflector. In acoustics, theoretical work has shown that Luneburg lens can focus the incoming wave on the opposite surface of the lens [11, 12]. Other gradient-index lenses have also been proposed to achieve acoustic focusing [13], asymmetric transmission [14], sound absorption [15] and so on. They have less stringent refractive index requirement than Eaton lenses and thus potentially allows for broadband operation. Experiments have also been done using pillar-type structures [16, 17], crossbuck geometries [18] and space-coiling metamaterials [19, 20] for the realization of gradient-index lenses. However, these approaches are either bulky or have limited working bandwidth. In this paper, the discretized refractive index profile of Luneburg lens is achieved with spiral unit cells with tapered geometry [21-23]. The structure is non-resonant and has better impedance matching compared with pillar type structures or crossbuck geometies. By adding a reflected boundary on the focal surface of a Luneburg lens, waves can be directed back along the incident direction (as illustrated in Fig. 1(a)) with minimal lateral shift.

---

[*] These authors contributed equally to this work
[†] kenyon@xmu.edu.cn
[‡] cummer@ee.duke.edu

Compared with Eaton lens-based or metasurface-based retroreflectors [7, 24], the scheme by combining a Luneburg lens and an arcual mirror releases the requirement of high index profile or complicated geometrical configurations.

To demonstrate the design concept, a compact acoustic retroflector is realized with a radius of only approximately one wavelength. The geometry-dependent properties also allow the device to be conveniently scaled to function at various frequencies including airborne ultrasound. Both the numerical and the experimental results show our designed acoustic retroreflector can function within a relatively wide incident angle range (from –60 degrees to +60 degrees) and over a bandwidth of about 27% of the central frequency.

## II. THEORETICAL MODEL AND NUMERICAL VERIFICATION

The schematic of the acoustic retroreflector is shown in Fig. 1(a), where the retroreflector is composed of two elements: a Luneburg lens (the refractive index is marked with a color gradient) and an arcual mirror (marked with gray color). The working principle of the acoustic retroreflector is displayed in Fig. 1(a), where the red and blue curves reveal the propagation trajectory of the incident and reflected waves in the acoustic retroreflector. Waves incident from the same direction (e.g., see the red solid and dashed curves in Fig. 1(a)) will converge at the same point locating at the surface of Luneburg lens, thanks to the refractive index profile of Luneburg lens given as $n = \sqrt{2-(r/R_0)^2}$, where $R_0$ is the radius of Luneburg lens. As the arcual area with hard materials can function as a mirror, the converged incident waves will be reflected and return symmetrically back through Luneburg lens. Consequently, the reflected waves will propagate along the direction that is parallel and opposite to the incident one (see the blue solid and dashed curves in Fig. 1(a)). In a pulse-echo based detection system, such retroreflection can significantly increase the signal-to-noise ratio of the echo signal and thus improve the sensing performance (e.g., longer detection range).

To verify this concept, the corresponding numerical result based on COMSOL Multiphysics is shown in Fig. 1(b), where the proposed acoustic retroreflector is comprised of an ideal Luneburg lens ($R_0 = 0.1m$) and an arcual ring (its central angle is $150°$, for the discussion of the central angle, please see FIG. S1 in Supplemental Material) of a sound-hard material. For the incident beam with an incident angle of $\theta = -30°$ (the incident angle is defined as the deviation angle from the geometric symmetric axis of the retroreflector, with the left deviation angle defined as "–"), there is enhanced pressure field in the incident area where the incident wave can enter into Luneburg lens (see the white dashed frame), implying that the propagation direction of the reflected wave (the blue arrow) is parallel and opposite to the incident one (the red arrow). To display the effect of acoustic retroreflector more clearly, the acoustic fields of the incident beam toward the retroreflector and the reflected beam returning from retroreflector are shown respectively in Fig. 1(c). From the information in Fig. 1(c), the reflected beam with planar wavefront, indeed, propagates along the direction parallel and opposite to the incident one, yet leading to some scattering waves (see the reflected field outside the white frame) caused by the wide incident beam toward the retroreflector (see the incident field outside the white frame). In fact, as the incident beam can converge at the interface between the Luneburg lens and the arcual area for a wide incident angle, accordingly such acoustic retroreflector can function in a wide incident angular range. Based on the above analysis and numerical results in Fig. 1, the wide angle acoustic retroreflector could be obtained by using a mirrored Luneburg lens.

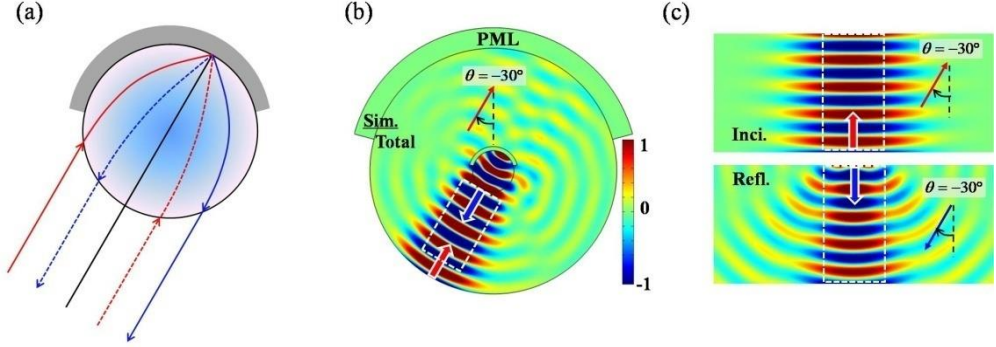

FIG. 1. (a) The schematic of an acoustic retroreflector composed of a Luneburg lens (gradient color) and an arcual reflector (gray color). In plot, these curves are the propagation trajectory of the acoustic retroreflector. (b) The simulated total acoustic pressure field of an ideal acoustic retroreflector with a radius of $R_0 = 0.1m$ for the incident beam with $\theta = -30°$. The working frequency is 3.0 kHz and PML is the perfect matched layer. (c) The corresponding simulated acoustic pressure field patterns of the incident beam (the red arrow) toward the retroreflector and the reflected beam from the retroreflector (the blue arrow).

## III. DESIGN AND FABRICATION

To discretize such a continuous gradient index profile required, the designed Luneburg lens with $R_0 = 0.1m$ is divided into five layers: a core with a radius of $r_0 = 0.02m$ and four circular rings with thicknesses of $0.02m$. The effective refractive index of each layer is given as $n_l = \sqrt{2 - ((l - 0.5)r_0 / R_0)^2}$ ($l$ denotes the order of the divided layers). Each layer is composed of unit cells with the effective refractive indices realized by acoustic metamaterials. Therefore, five different kinds of acoustic metamaterial are designed for these unit cells. The core region is filled with a single unit cell and its size is $0.04m \times 0.04m$. The unit cells in the four circular ring regions are the same sizes, i.e., $0.02m \times 0.02m$, and as a result there are successively 6, 12, 24, 36 unit cells from the inner circular ring to the outer circular ring. Although many topologies can be adopted for the unit cell design with discrete refractive index, here we choose tapered spiral cells as the candidate. Compared with structures such as pillars [16, 17], crossbuck structure [18] or perforated plates [25, 26], the spiral cells we used here have simple geometrical configurations and better impedance matching, and it is expected to have relatively low loss. To illustrate the design details of acoustic metamaterials in these unit cells, here we take the unit cell for obtaining $n_3 = 1.33$ ($l=3$) as an example shown in Fig. 2(a), where the unit cell is composed of four identical microstructures possessing four-fold rotational symmetry. The thickness of the wall is $t = 0.001m$, with its geometrical shape described by the Archimedean spiral [21]. By designing the appropriate parameters for the Archimedean spiral, the required effective refractive index of this unitcell can be obtained. The effective refractive index is retrieved from the transmission and reflection coefficients of the unit cell in a waveguide [27] (see the simulated field pattern in Fig. 2(a)). Accordingly, by designing five different kinds of acoustic metamaterials, the Luneburg lens working at 3 kHz was designed shown in Fig. 2(b), where these different acoustic metamaterials are respectively placed in the corresponding areas to realize these particular indices marked by different colors. Although these designed unit cells have slight geometric deformation with the curved sectors and the central circle, the sizes of the unit cells are much smaller compared with the working wavelength, and therefore the geometric deformations are assumed to have negligible effect on the effective indices. To show the dispersion of the unit cells, the effective refractive index is also plotted as a function of frequency, as depicted in Fig. 2(c). Within the frequency range from 2.5 kHz to 3.5 kHz, the effective indices remain roughly the same, which indicates that the unit cells have broadband performance.

Furthermore, based on the design of the Luneburg lens, the acoustic retroreflector working at 3.0 kHz is fabricated by using fused filament 3D printing technology (see inset of Fig. 2(d)). To test the performance of the acoustic retroreflector, the experimental setup is schematically shown in Fig. 2(d). The incident wave is a Gaussian modulated plane wave generated by the speaker array, and the acoustic field is scanned with a moving microphone. To measure the sample response at different incident angles, the sample is rotated while the incident beam is unchanged.

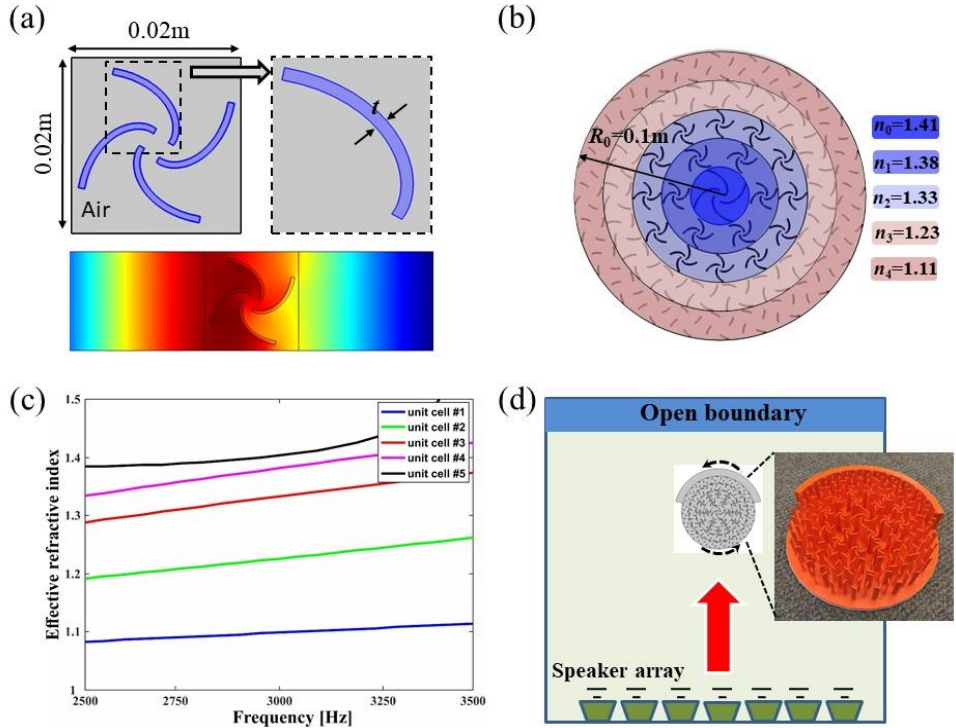

FIG. 2. (a) The design details of the spiral acoustic metamaterials. (b) The designed Luneburg lens working at 3.0 kHz with discrete index profiles realized by gradient acoustic metamaterials. (c) Effective refractive index of the unit cells as a function of frequency. (d) The schematic of the experimental setup for the acoustic retroreflector. Inset shows the fabricated simple of the acoustic retroreflector working at 3.0 kHz.

## IV. EXPERIMENTAL DEMONSTRATION
### A. Near field performance

Figures 3(a) and 3(b) show the simulated total pressure field and the simulated and measured reflected field for a beam incident on the designed retroreflector at $\theta = 0°$. From the simulation in Fig. 3(a), the total pressure field is enhanced in the incident area (see the white dashed frame), which means that the reflected wave (the blue arrow) propagates towards the incident one (the red arrow). To observe the performance of the designed acoustic retroreflector clearly, the simulated reflected pressure field is shown in the left part of Fig. 3(b), where a flattened wavefront of a Luneburg lens is seen in the retroreflector. Particularly, the reflected beam in the incident area marked by the white dashed frame possesses a planar wavefront, implying that the incident beam is retroreflected and not broadly scattered.

The performance of the retroreflector is then verified experimentally using the 3D printed sample. The retroreflector is situated in a two-dimentional waveguide. The incident wave is a Gaussian modulated plane wave generated by the speaker array with its propagation direction normal to the flat speaker array. The acoustic field is scanned with a moving microphone with a step size of 2cm. The received signals are then time windowed to extract the incident and reflected waves. Fourier transform is then performed to obtain the acoustic field information at different frequencies. To measure the sample response at different incident angles, the sample is rotated at the corresponding angles while keeping the incident beam unchanged. The corresponding measured reflected pressure field, marked by the black dashed frame is shown in the right side of Fig. 3(b), shows that the experimental result is in good agreement with the simulated one. Although the wavefront curvature outside of the incident area caused by arcual reflector, the wavefront of the reflected field in the incident area, indeed, is planar. To highlight this feature of our designed retroreflector, the corresponding results of an incident beam with $\theta = 0°$ on a cylinder with $R_0 = 0.1m$ are shown in Figs. 3(c) and 3(d). Although there is enhanced total pressure field in the incident area, yet with more scattering wave (see Fig. 3(c)). However, by observing the simulated reflected field pattern in the left image of Fig. 3(d), there is a monopole-like, omnidirectional cylindrical radiation. This is confirmed by the measured result in the right plot of Fig. 3(d). By comparing the simulated reflected fields in Fig. 3, the reflected field from the retroreflector is stronger than that from a cylinder. It is because that, concerning the property of the reflected wave in an ideal case, the reflected wave from the proposed retroreflector in the incident area is unattenuated radiation with planar wavefront, while it is attenuated radiation with cylindrical wavefront for the reflected wave from a cylinder. But from both measured reflected results in Fig. 3, the reflected field from the retroreflector is not stronger than that from the cylinder, which is caused by loss in the designed device. If the reflected wave is measured in far field, the enhanced reflected wave with planar wavefront should be received from the retroreflector, in comparison to that from a cylinder.

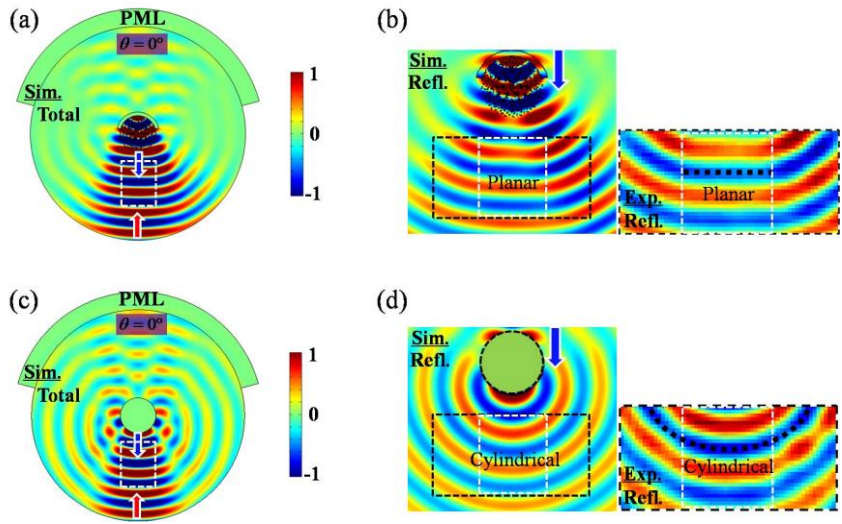

FIG. 3. (a) and (b) are the simulated total pressure field pattern and the simulated (left) and the measured (right) reflected pressure field patterns from retroreflector for an incident beam at $\theta = 0°$ on the designed retroreflector at 3.0 kHz. (c) and (d) are the simulated total

pressure field pattern and the simulated (left) and the measured (right) reflected pressure field patterns for an incident beam with $\theta = 0°$ on a cylinder with $R_0 = 0.1m$.

Similar retroreflection performance is seen in Fig. 4 for the incident angles with $\theta = -30°$ and $\theta = -60°$. For $\theta = -30°$, the enhanced total pressure field is seen in Fig. 4 (a) and the simulated and measured reflected fields consistently reveal the planar wavefront in the incident area (see Fig. 4(b)). While for $\theta = -60°$, as a part of the focusing region of the incident wave is beyond the arcual reflector, some incident wave penetrates through the retroreflector (see Fig. 4(c)). As a result, the intensity of the reflected wave is reduced and the performance of planar wavefront is decreased caused by more scattering from the arcual reflector (see Fig. 4(d)). As the designed acoustic retroreflector possesses a symmetric index profile, the designed acoustic retroreflector can function for a wide incident angle, with a range from $\theta = -60°$ to $\theta = 60°$. Accordingly, wide angle acoustic retroreflector performance at 3.0 kHz is clearly demonstrated in both simulated and experimental results. In addition, the wide angle designed acoustic retroreflector also performs well at other frequencies, as shown in the corresponding simulated and measured reflected waves at 2.6 kHz and 3.4 kHz (see FIG. S2 in Supplemental Material).

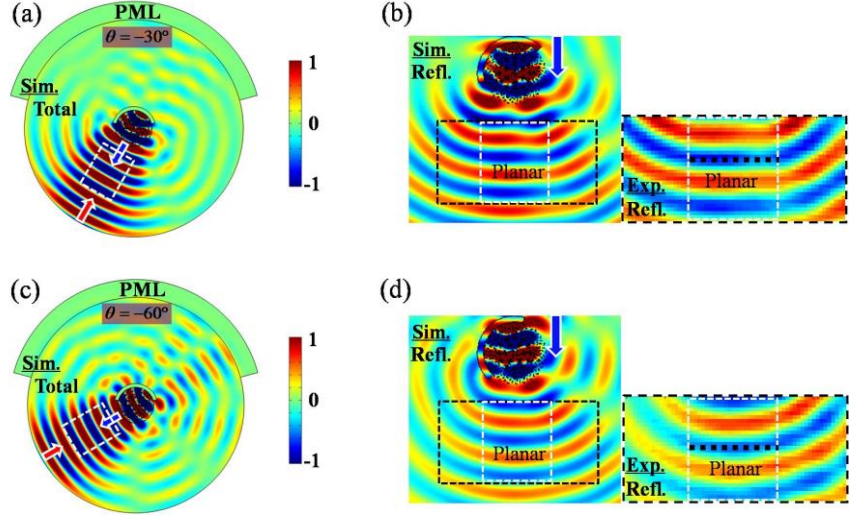

FIG. 4. (a) and (b) are the simulated total pressure field pattern and the simulated (left) and the measured (right) reflected pressure field patterns from the retroreflector for the incident beams with $\theta = -30°$ bumping on the designed retroreflector at 3.0 kHz, respectively. (c) and (d) are the simulated total pressure field pattern and the simulated (left) and the measured (right) reflected pressure field patterns from retroreflector for the incident beams with $\theta = -60°$ bumping on the retroreflector at 3.0 kHz, respectively.

## B. Far field performance

We now extrapolate from the near-field measurements and simulations to determine the far-field performance of this acoustic retroreflector. As an illustration, the reflected signal in the far field for normal incidence ($\theta = 0°$) is shown in Fig. 5, where the normalized reflected power at a distance of $D = 50\lambda$ is numerically and experimentally revealed in a polar pattern. In both the numerical and experimental plots, the reflected power at $D = 50\lambda$ in the direction of $0°$ for a beam incident on a $R_0 = 0.1m$ cylinder with $\theta = 0°$ is normalized as unity.

We first focus on the numerical results. From Fig. 5(a), the reflected energy from the cylinder is almost uniform in all directions, caused by the cylindrical radiation (see the black dotted curve). In contrast, for the designed retroreflector at 3.0 kHz (see the red solid curve in Fig. 5(a)), the highest reflected power is observed in the direction of 0°. For this case the retroreflected power from the retroreflector is four times that from the cylinder. There is enhanced power in the 30° direction in the simulation (see the red solid curve in Fig. 5(a)), and also a weak enhancement in the –30° direction. Simulations of the reflection at normal incidence from the arcual reflector by itself (see the gray dashed curve in Fig. 5(a)) shows significant reflection –30° and 30°. We thus attribute the reflected power in these two directions to the arcual reflector of the retroreflector. The asymmetry of reflection in these two directions is caused by the spiral microstructures in the retroreflector.

The experimental measurements confirm the good expected retroreflector performance. The experimental result for a normally incident beam on the retroreflector at 3.0 kHz (see the red dot-dashed curve) is also shown in Fig. 5(a). Here the experimental data in far field are obtained through the projection of the measured near field based on Huygens' principle. There is less reflected power in the direction of 0° in experiment caused by the loss in the designed retroreflector, but the experimental and numerical results are in good agreement. Relative to the retroreflected power from a cylinder, there is nearly three times more retroreflected power from the designed retroreflector in experiment.

We performed simulations for normal incidence from 2.0 to 4.0 kHz to assess the device bandwidth. Performance is expected to drop at lower frequencies because the device becomes smaller relative to a wavelength, and also at higher frequencies because the discretized Luneburg approximation begins to break down. We find that good retroreflection performance extends to the range between 2.6 kHz and 3.4 kHz, which are shown by the blue and green curves in Fig. 5 (b). The experimental results agree well with the simulations. Therefore, the enhanced and directional reflection is well demonstrated in our designed acoustic retroreflector, functioning in a 27% bandwidth about the central design frequency.

As the designed retroreflector can fucntion in a wide incident range, the similar results can be achieved for other incident angles, e.g., see the case of $\theta = -30°$ in FIG. S3 in Supplemental Material. In fact, even the retroreflector is not placed in the central direction of the incident beam, e.g., it deviates from the central axis with a distance of one wavelength, the enhanced retroreflected signal can be observed as well (see FIG. S4 in Supplemental Material). It is also worth noting that our retroreflector's radius is approximately equal to one wavelength. However, if the size of our proposed retroreflector is increased, the enhanced and directional reflection will be improved further (see FIG. S5 in Supplemental Material).

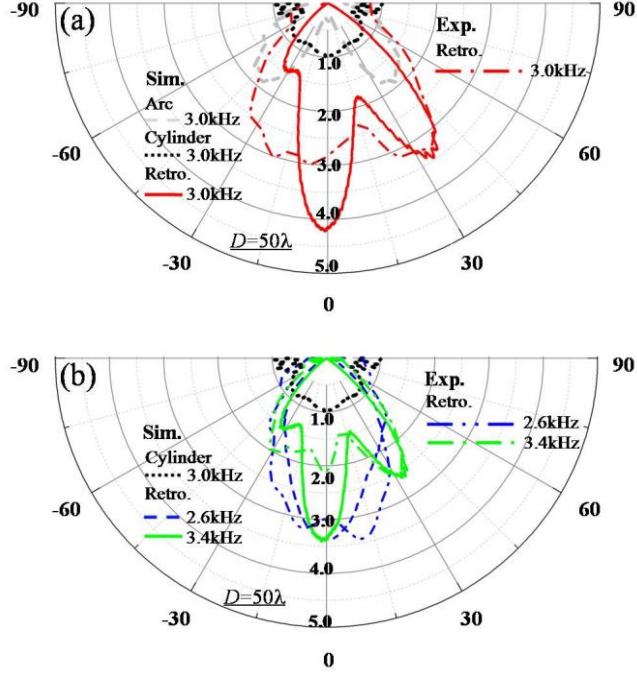

FIG. 5. (a) The polar pattern of the normalized reflected power at $D=50\lambda$ for a normally incident beam ($\theta=0°$) on different structures. The gray dashed and black dotted curves are the cases of the reflected power from the arcual ring and cylinder at 3.0 kHz respectively, and the red solid and red dot-dashed curves are the cases of the numerically and expermentally reflected power from the designed retroreflector at 3.0 kHz, respectively. (b) The polar pattern of the normalized reflected power at $D=50\lambda$ for the normally incident beam ($\theta=0°$) on the designed retroreflector at other frequencies. The blue dashed (green solid) and blue dot-dot-dashed (green dot-dashed) curves are the cases of the numerically and expermentally reflected power from the designed retroreflector at 2.6 kHz (3.4 kHz), respectively.

## V. CONCLUSIONS

In conclusion, we have designed a compact acoustic mirrored Luneburg lens by placing a rigid arcual back ajacent to the original lens. The device greatly expands the functionality of the Luneburg lens and can perform as an acoustic retroreflector. The retroflector, with a radius of only approximately one wavelength, works in an incident angular range up to 120 degrees over a relatively broad bandwidth of about 27% of the central frequency. In principle, the design can be extended to the airborne ultrasound frequency range and three dimensions [28, 29] and the performance of the proposed retroreflector might be well optimized using other acoustic metamaterials [30, 31]. In addition, we have also performed transient simulations to study the pule-echo based performance of the retroreflector [see Supplemental Material Videos 1, 2 and 3 for the full time evolution of the pressure fields], and the results are consistent with the the frequency domain simulation. The detailed setup and results are summarized in Fig. S6 in Supplemental Material. Therefore, we expect our proposed retroreflector to be potentially useful for pulse-echo-based acoustic sensing and communication, and is expected to be of interest to researchers in the areas such as robotic sensing, ultrasonic imaging, distributed sensor networks.

**Acknowledgement**

Y. F. would like to acknowledge the Chinese Scholarship Council (No.201606920045). J. L., Y. X., C. S., S. A. C. were supported by a Multidisciplinary University Research Initiative grant from the Office of Naval Research (N00014–13-1–0631) and an Emerging Frontiers in Research and Innovation grant from the US National Science Foundation (grant 16-41084). Y. X. was supported by National Natural Science Foundation of China (NSFC) (11604229), and H. C. was supported by the National Science Foundation of China for Excellent Young Scientists (Grant no. 61322504).